# Procedure of tuning up a three-site artificial Kitaev chain based on transmon measurements

Xiaozhou Yang[1,2,†], Zhaozheng Lyu[1,2,3,†,*], Xiang Wang[1,2], Enna Zhuo[1,2], Yunxiao Zhang[1,2], Duolin Wang[1,2], Yukun Shi[1,2], Yuyang Huang[1,2], Bing Li[1,2], Xiaohui Song[1,2,4], Peiling Li[1], Bingbing Tong[1], Ziwei Dou[1], Jie Shen[1,2,4], Guangtong Liu[1,2,3,4], Fanming Qu[1,2,3,4], and Li Lu[1,2,3,4,*]

[1] *Beijing National Laboratory for Condensed Matter Physics, Institute of Physics, Chinese Academy of Sciences, Beijing 100190, China*

[2] *School of Physical Sciences, University of Chinese Academy of Sciences, Beijing 100049, China*

[3] *Hefei National Laboratory, Hefei 230088, China*

[4] *Songshan Lake Materials Laboratory, Dongguan, Guangdong 523808, China*

[†] These authors contribute equally to this work.

[*] Corresponding authors: Zhaozheng Lyu: lyuzhzh@iphy.ac.cn, Li Lu: lilu@iphy.ac.cn

**Abstract**

Artificial Kitaev chains (AKCs), formed of quantum dot-superconductor linear arrays, provide a promising platform for hosting Majorana bound states (MBSs) and implementing topological quantum computing. The main challenges along this research direction would include the tuning up of AKCs for hosting MBSs and the readout of the parity of the chains. In this work, we present a step-by-step procedure for tuning up a three-site AKC to its sweet spots based on the spectra of a transmon circuit which is integrated with the chain for the purpose of reading out the parity of the chain. The signatures of the transmon's plasma modes in each step, particular those related to the appearance of MBSs in the chain, will be given. We find that the sweet spots in a three-site AKC can be classified into three types based on the relative strengths of elastic cotunneling (ECT) and crossed Andreev reflection (CAR): ECT-dominated sweet spots, genuine sweet spots and CAR-dominated sweet spots. We show that the ECT-dominated and CAR-dominated sweet spots can be more conveniently accessed and utilized in transmon-based measurements.

## I. Introduction

MBSs in solid-state systems can be used to form topological qubits due to their non-Abelian exchange statistics [1–3]. Various platforms have been proposed and investigated as candidates for hosting MBSs [4–10]. The simplest theoretical model among them is the one-dimensional spinless fermion chain with p-wave superconducting pairing, known as the Kitaev chain [4]. In practical implementations, it has been proposed to build an artificial Kitaev chain site-by-site via an array of quantum dot (QD) and superconductor (S) sites [11,12]. In this proposal, spin-polarized quantum dots act as spin-less fermions. They are coupled with each other via ECT and CAR processes through the superconductor between them [13,14]. It has been predicted that even a chain with very few sites can host a pair of zero-energy bound states at the sweet spots, known as the poor man's Majorana bound states, if the ECT and CAR strengths can be fine-tuned [15–20]. In experiments, conductance signatures consistent with MBSs have been detected in both two-site and three-site chains fabricated on nanowires and two-dimensional electron gases [21–27]. These progresses have inspired numerous studies on the optimization of the AKC's configurations [28–30], the scaling up of the systems [19,31–33], and on how to demonstrate the non-Abelian statistics [34–37] and to construct the Majorana qubits [36–39], just to mention a few.

Detecting MBSs by circuit quantum electrodynamics (cQED) measurements is one of the key steps in constructing Majorana qubits. Recently, the fermion parity of a two-site chain has been read out using a transmon circuit, and the simultaneous detection of odd-even degenerate ground states has been achieved [40]. However, these measurements were not performed at the sweet spots. In order to find the sweet spots and reveal the signature of MBSs at the sweet spot in transmon-based measurements, in this study we present a procedure that can be followed experimentally for a three-site AKC integrated into a transmon circuit.

## II. Model Device and Hamiltonian

We consider a three-site AKC integrated into a transmon circuit, as shown in Fig. 1. The chain is modeled as a QD-S linear array, consisting of three quantum dots and two superconducting dots arranged alternately, as described in Refs. [19,20]. The superconducting dots are connected in parallel with a reference Josephson junction to form a superconducting quantum interference device (SQUID).

First, let us focus on the three-site AKC. The Hamiltonian is

$$H_{\text{AKC}} = \sum_{j=1,3,5} H_{\text{QD}}^j + \sum_{j=2,4} H_{\text{S}}^j + \sum_{j=1}^{4} H_{\text{T}}^j,$$

$$H_{\text{QD}}^j = \sum_{\sigma=\uparrow,\downarrow} (\varepsilon_j + \eta_\sigma E_Z) n_{j,\sigma} + U n_{j,\uparrow} n_{j,\downarrow},$$

$$H_{\text{S}}^j = \sum_{\sigma=\uparrow,\downarrow} \varepsilon_j n_{j,\sigma} + \left(\Delta e^{i\phi_j} c_{j,\uparrow}^\dagger c_{j,\downarrow}^\dagger + \text{h.c.}\right),$$

$$H_{\text{T}}^j = t_j \sum_{\sigma=\uparrow,\downarrow} \left(t_{\text{sc}} c_{j,\sigma}^\dagger c_{j+1,\sigma} + \eta_\sigma t_{\text{sf}} c_{j,\sigma}^\dagger c_{j+1,\bar\sigma} + \text{h.c.}\right), \qquad (1)$$

where $n_{j,\sigma} = c_{j,\sigma}^\dagger c_{j,\sigma}$, $c_{j,\sigma}^\dagger$ ($c_{j,\sigma}$) is the electron creation (annihilation) operator and $\bar\sigma$ means the opposite spin to $\sigma$. $H_{\text{QD}}^j$ is the Hamiltonian of the quantum dot with the odd index $j$, $\varepsilon_j$ is the on-site energy, $\eta_\uparrow = +1$ ($\eta_\downarrow = -1$), $E_Z$ denotes the Zeeman spin splitting and $U$ represents the

Coulomb repulsion. $H_S^j$ is the Hamiltonian of the superconducting dot with the even index $j$, $\Delta$ denotes the induced superconducting pairing potential, and $\phi_j$ represents the phase. $H_T^j$ is the nearest-neighbor coupling, $t_j$ denotes the coupling strength, and $t_{sc}$ ($t_{sf}$) represents the spin-conserving (spin-flipping) component. In the rest of this work, we choose $\Delta$ as the energy unit and $\phi_2 = 0$. Unless stated otherwise, we set $E_Z = 3\Delta$, $U = 5\Delta$, $\phi_4 = 0$, $t_j = 0.4\Delta$, $t_{sc}/t_{sf} = 3$ and $t_{sc}^2 + t_{sf}^2 = 1$, according to recent works [19,31].

To identify the sweet spots in an AKC, we introduce two key quantities [18,32,41]. The first is the energy splitting between odd and even parity ground states:

$$\delta E_{00}^{oe} = E_0^o - E_0^e, \tag{2}$$

where the superscript of $E_m^{o/e}$ represents the odd/even parity and the subscript indicates the eigenstate quantum number (with $m = 0$ corresponding to the lowest energy state). The existence of MBSs predicts a near-zero $\delta E_{00}^{oe}$. The second quantity is the Majorana polarization at dot $j$:

$$MP_j = \frac{\sum_{\sigma=\uparrow,\downarrow}(\omega_{j,\sigma}^2 - z_{j,\sigma}^2)}{\sum_{\sigma=\uparrow,\downarrow}(\omega_{j,\sigma}^2 + z_{j,\sigma}^2)}, \tag{3}$$

where $\omega_{j,\sigma} = \langle o|(c_{j,\sigma} + c_{j,\sigma}^\dagger)|e\rangle$ and $z_{j,\sigma} = \langle o|(c_{j,\sigma} - c_{j,\sigma}^\dagger)|e\rangle$. A near-one $|MP_j|$ indicates one MBS is located at dot $j$. Additionally, two other quantities are calculated to discuss the properties of sweet spots [20,29,32]: the excitation gap

$$\delta E_{gap} = \min(\delta E_{10}^{oo}, \delta E_{10}^{ee}), \tag{4}$$

and the Majorana wavefunctions

$$\psi_j^{A/B} = \sum_{\sigma=\uparrow,\downarrow} |\langle o|(c_{j,\sigma} \pm c_{j,\sigma}^\dagger)|e\rangle|^2. \tag{5}$$

Second, let us focus on the plasma modes of the transmon circuit. In the SQUID loop, the phase difference across the chain $\phi = \phi_4 - \phi_2$ and the reference junction satisfy the equation: $\phi = \theta + \phi_{ext}$, where $\theta$ is the phase difference across the reference junction, and $\phi_{ext}$ is the phase difference applied by an external magnetic flux. In the model device we designed, the Josephson energy $E_J$ of the reference junction is significantly larger than that of the chain, so $\theta$ is near 0 and $\phi$ is close to $\phi_{ext}$. In this case, the transmon Hamiltonian is:

$$H = -4E_C \partial_\theta^2 - E_J \cos(\theta) + H_{AKC}(\theta + \phi_{ext}), \tag{6}$$

where $\partial_\theta^2$ denotes the second partial derivative with respect to $\theta$, $E_C$ represents the charging energy of the transmon circuit which satisfies the transmon limit $E_J \gg E_C$. The energy difference between the two lowest transmon states $\delta E_p$ can be derived using second-order perturbation theory. By considering the three lowest unperturbed transmon states, neglecting the transitions between chain eigenstates, and retaining terms up to $\lambda^{-2}$, the expression can be written as

$$\delta E_p = (\hbar\omega_p - E_C) + \frac{1}{2\lambda^2} \partial_\phi^2 E_{AKC}|_{\phi=\phi_{ext}}, \tag{7}$$

where $\hbar\omega_p = \sqrt{8E_J E_C}$, $\lambda = (E_J/8E_C)^{1/4}$, and $\partial_\phi^2$ denotes the second partial derivative with respect to $\phi$ (see Supplemental Materials for detailed derivations). Since the first term and the coefficient of the second term are constants, we concentrate on $\partial_\phi^2 E_{AKC}$ in the following analysis.

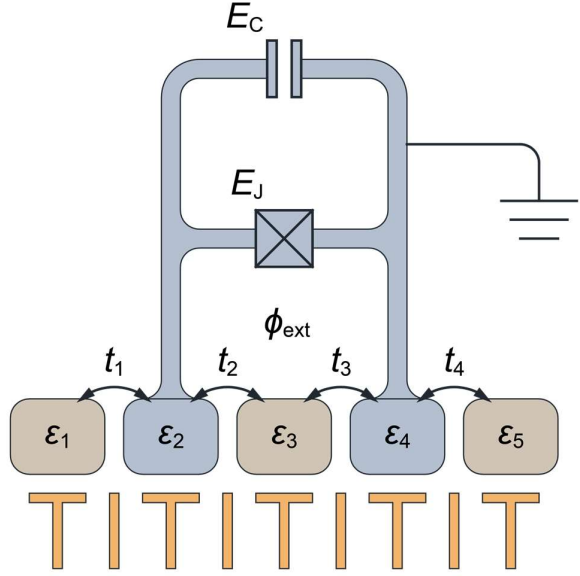

**Fig. 1.** Schematic illustration of a three-site AKC integrated into a transmon circuit with $E_J$ and $E_C$. The chain consists of three quantum dots (brown) and two superconducting dots (blue). The on-site energy $\varepsilon_i$ and the inter-dot coupling $t_i$ can be tuned by the gate electrodes (yellow). The phase difference across the superconducting dots can be controlled by an external magnetic flux $\phi_{\text{ext}}$.

### III. Procedure Details

In our model device, three different configurations can be implemented step by step through gate manipulations. The simplest configuration allowing to couple with a transmon circuit is the S-QD-S configuration. Let us first discuss this configuration (step 1), then expand to subsequent configurations of QD-S-QD-S (step 2) and QD-S-QD-S-QD (step 3, a three-site AKC).

*Step 1.* The S-QD-S configuration can be realized by gate manipulations to set $t_1 = t_4 = 0$, as shown in Fig. 2(a). This step aims to look for odd-even parity transition in the quantum dot (dot 3) and near-zero on-site energies in the superconducting dots (dots 2 and 4). First, we can identify the parity of ground state through the sign of the second derivative of the ground state energy $\partial_\phi^2 E_0$: it exhibits a negative value in the odd parity ground state and a positive value in the even parity ground state, with a discontinuity at the parity boundary, as shown in Fig. 2(b). The weak inter-dot coupling favors the odd parity ground state, as shown in the parity phase diagram in Fig. 2(c). These results are in agreement with previous findings in Ref. [42].

Since the Andreev bound states of the superconducting dots cannot be directly observed in transmon-based measurements, we need to analyze the coupling diagrams between the quantum dot and the superconducting dots. Figures 2 (d)-(f) show $\partial_\phi^2 E_0$ as a function of $\varepsilon_3$ and $\varepsilon_{2,4}$ at different coupling strengths. In the weak coupling regime, the parity boundary exhibits the largest distortion when $\varepsilon_{2,4}$ approaches zero. Meanwhile, $\partial_\phi^2 E_0$ reaches its minimum for a given $\varepsilon_3$ in the odd parity region (or the maximum in the even parity region), as shown in Fig. 2(e) and (f). However, it is difficult to maintain $\varepsilon_2 = \varepsilon_4$ in experiments. We then demonstrate that the features of $\partial_\phi^2 E_0$ persist without this condition, by scanning $\varepsilon_2$ and fixing $\varepsilon_4$. Figure 2(g) shows the extracted $\varepsilon_{2,\text{ex}}$ (denoting the $\varepsilon_2$ at the $\partial_\phi^2 E_0$ minimum) as a function of $\varepsilon_4$ at different coupling strengths while setting $\varepsilon_3 = -2.5\Delta$. Although the maximum deviation of $\varepsilon_{2,\text{ex}}$ from zero increases with the coupling strengths, overall its value is very close to zero, not exceeding $0.02\Delta$ even when $t_2 = t_3 = 1.0\Delta$. This provides us a feasible way to tune dot 2 to the near-zero $\varepsilon_2$ from the coupling diagrams. The same procedure can also be applied to tune dot 4 to $\varepsilon_4 \approx 0$. Experimentally, we have observed a coupling diagram similar to Fig. 2(e) in cQED measurements (see Supplementary Information for details). Meanwhile, the conductance and supercurrent signatures in Ref. [43] have similar features.

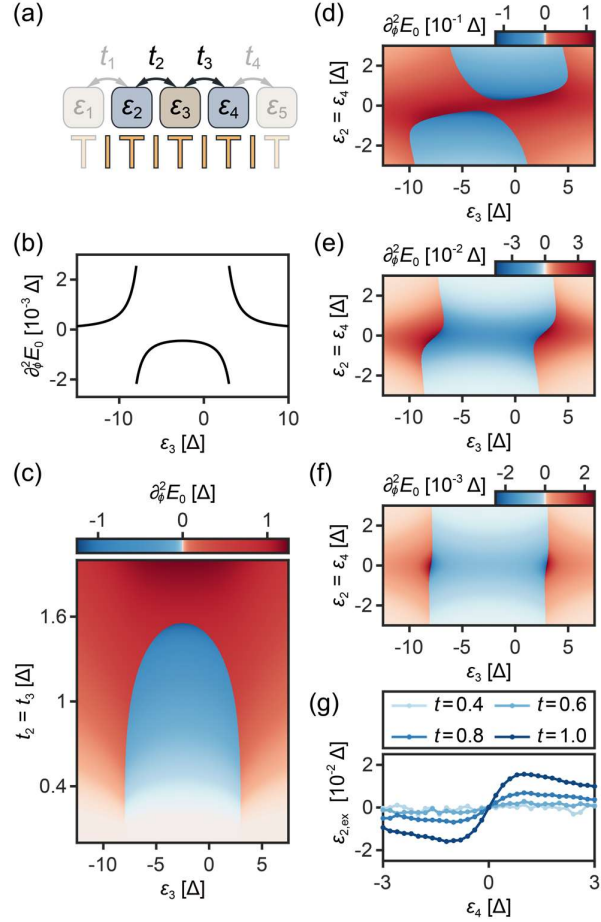

**FIG. 2.** (a) Schematic illustration of the QD-S-QD configuration, with dots 1 and 5 decoupled ($t_1 = t_4 = 0$). (b) The second derivative of the ground state energy $\partial_\phi^2 E_0$ as a function of $\varepsilon_3$, with settings $t_2 = t_3 = 0.4\,\Delta$. (c) The phase diagram: $\partial_\phi^2 E_0$ as a function of $\varepsilon_3$ and $t_{2,3}$. The positive and negative values corresponding to even and odd parity states. (d-f) The coupling diagrams: $\partial_\phi^2 E_0$ as a function of $\varepsilon_3$ and $\varepsilon_{2,4}$ at coupling strengths of $t_2 = t_3 = 1.6\,\Delta$, $1.0\,\Delta$ and $0.4\,\Delta$, respectively. (g) The extracted $\varepsilon_{2,\text{ex}}$ as a function of $\varepsilon_4$, with fixing $\varepsilon_3 = -0.25\,\Delta$. The coupling strengths are set to $t_2 = t_3 = 0.4\,\Delta$, $0.6\,\Delta$, $0.8\,\Delta$ and $1.0\,\Delta$, with the corresponding colors gradually changing from light to dark.

*Step 2.* We then consider the QD-S-QD-S configuration by activating dot 1, as shown in Fig. 3(a). For this configuration, odd-even parity transition in dot 1 has been observed in a recent experiment [40]. Here, we focus on the identification of sweet spots in this configuration. We keep the parameters of the right-hand S-QD-S structure the same as those in Fig. 2(b), except for setting $\varepsilon_3 = 4\Delta$ to enhance the detection sensitivity for dot 1. By varying $\varepsilon_3$ and $t_1$, we find that $\partial_\phi^2 E_0$ does not exhibit a sign difference between the odd and even parity ground states, but shows a dip at their boundary, as shown in Fig. 3(b). Moreover, the parity boundary in Fig. 3(c) exhibits a subtle leftward tilt in comparison to that in Fig. 2(c). A sweet spot in the QD-S-QD-S configuration requires equal strengths of ECT and CAR between two adjacent quantum dots. A continuous variation of $\varepsilon_2$ from zero to negative values can induce a transition from CAR dominance to ECT dominance across a sweet spot [13,14,19]. This process is manifested through the evolution of connectivity in Fig 3(d), (e) and (f), with settings $\varepsilon_2 = \varepsilon_{2,SP} - 0.3\Delta$, $\varepsilon_2 = \varepsilon_{2,SP}$ and $\varepsilon_2 = \varepsilon_{2,SP} + 0.3\Delta$, respectively. Here, $\varepsilon_{j,SP}$ denotes the value of $\varepsilon_j$ at the sweet spot. Additionally, $\varepsilon_{j,SP}$ ($j = 1,2,3$) is influenced by $\varepsilon_{4,SP}$. To quantify this relationship, we calculate $\varepsilon_{j,SP}$ as functions of $\varepsilon_4$, as shown in Fig. 3(g). The results demonstrate that the variation in $\varepsilon_4$ has a negligible impact on $\varepsilon_1$, ensuring that $\varepsilon_1$ can be fixed at $\varepsilon_{1,SP}$ once a sweet spot is identified. This discovery enables an iterative parameter optimization process that alternates between the QD-S-QD-S configuration (with dot 5 decoupled) and the S-QD-S-QD configuration (with dot 1 decoupled) to converge toward a special sweet spot. Each iteration alternately adjusts only two parameters, which is experimentally feasible. Here, we assume this special sweet spot corresponds to the dashed line in Fig. 3(g), where $\varepsilon_2 = \varepsilon_4 \approx -0.694\Delta$.

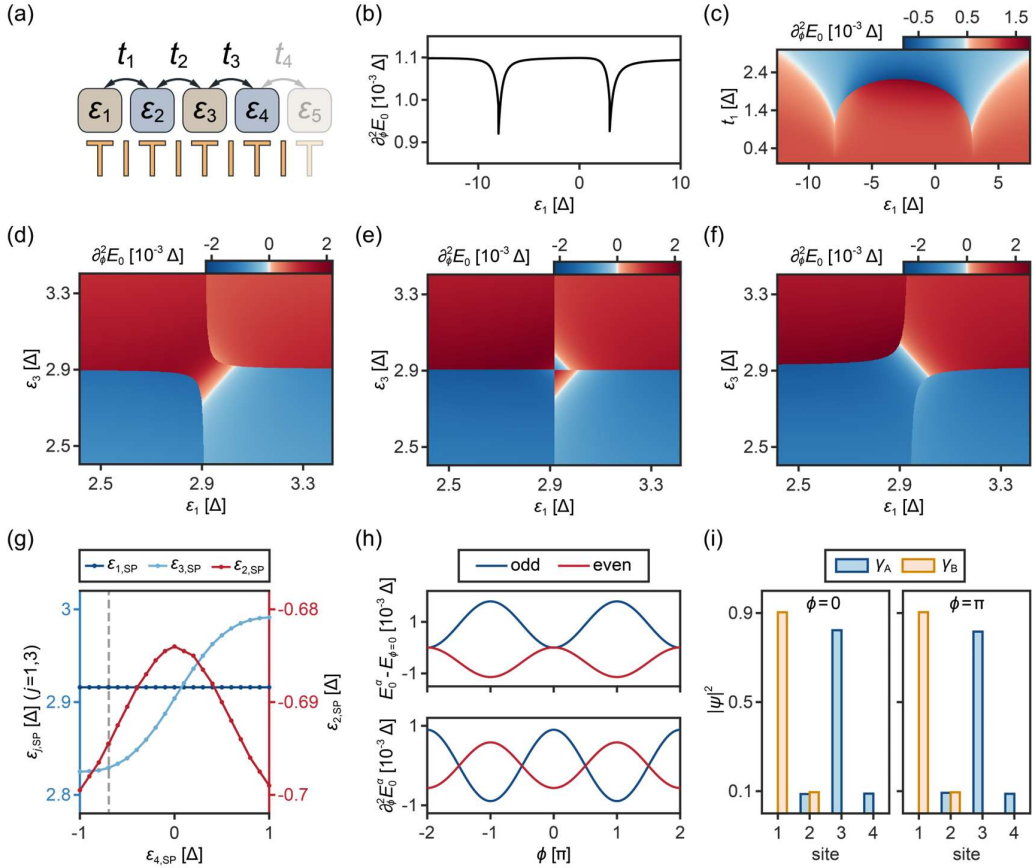

FIG. 3. (a) Schematic illustration of the QD-S-QD-S configuration, with dot 5 decoupled ($t_4 = 0$).

(b) $\partial_\phi^2 E_0$ as a function of $\varepsilon_1$, with setting $t_1 = 0.4\Delta$. (c) $\partial_\phi^2 E_0$ as a function of $\varepsilon_1$ and $t_1$. The dip corresponds to the odd-even parity boundary. The settings in (b) and (c) are $\varepsilon_2 = \varepsilon_4 = 0$, $\varepsilon_3 = 4.0\,\Delta$ and $t_2 = t_3 = 0.4\Delta$. (d), (e) and (f) $\partial_\phi^2 E_0$ as a function of $\varepsilon_1$ and $\varepsilon_3$ in the ECT-dominated regime, at the sweet spot and in the CAR-dominated regime, with setting $\varepsilon_2 = \varepsilon_{2,\mathrm{SP}} - 0.3\Delta$, $\varepsilon_2 = \varepsilon_{2,\mathrm{SP}}$, and $\varepsilon_2 = \varepsilon_{2,\mathrm{SP}} + 0.3\Delta$, respectively. Here, $\varepsilon_{2,\mathrm{SP}} = -0.684\Delta$ and $\varepsilon_{4,\mathrm{SP}} = 0$. (g) $\varepsilon_{j,\mathrm{SP}}$ ($j = 1,2,3$) as functions of $\varepsilon_{4,\mathrm{SP}}$. Regardless the value of $\varepsilon_4$, a sweet spot can always be realized. (h) The energy-phase relations of $E_0^o$ and $E_0^e$ in the top panel, and $\partial_\phi^2 E_0^{o/e}$ as functions of $\phi$ in the bottom panel. The spectrum reveals that the number of degenerate states is 2. (i) The Majorana wavefunction at the sweet spot, with $\phi = 0$ and $\pi$ in the left and right panel, respectively.

Next, we discuss the behavior of the sweet spot with the variation of phase difference. $\delta E_{00}^{oe}$ has a value less than $10^{-4}\Delta$ at $\phi = 0$ and gradually increases to $3 \times 10^{-3}\Delta$ as $\phi$ sweeps from 0 to $\pi$, as shown in the top panel of Fig. 3(h). If the ground states are degenerate within the energy scale of thermal fluctuations, their corresponding $\partial_\phi^2 E_0$ can be simultaneously observed [40], and $\partial_\phi^2 E_0^{o/e}$ are shown in the bottom panel of Fig.3(h). Conversely, we can determine the number of degenerate states through the transmon spectrum. Fig. 3(i) displays the Majorana wavefunctions at $\phi = 0$ and $\pi$, showing no significant differences between these two values of phase difference.

*Step 3.* We can identify a special sweet spot and prepare the value of dot 5 after the iterative process described in Step 2. By carefully checking the ECT and CAR behaviors near $\varepsilon_{2,\mathrm{SP}}$ in the QD-S-QD-S configuration and near $\varepsilon_{4,\mathrm{SP}}$ in the S-QD-S-QD configuration, we can ensure that $\varepsilon_{2,\mathrm{SP}}$ and $\varepsilon_{4,\mathrm{SP}}$ are of the same sign. Then, we can obtain a sweet spot in the QD-S-QD-S-QD configuration by activating dot 5. Conversely, if they are of the opposite sign, a reconfiguration process is required to identify a correct sweet spot. A systematic protocol for constructing a sign-ordered three-site AKC is proposed in Ref. [19]. We will not repeat the details here.

**IV. Sweet spots in the three-site AKC**

Although the sweet spots in three-site AKCs have been observed experimentally [23,26,27], further investigation in QD-S-QD-S-QD chains is still needed [19,20,44,45]. By assuming equal strengths of the ECT and CAR, a systematic analysis is performed considering the different on-site energies of quantum dots [20]. However, MBSs exist in a larger parametric region [16,26]. Here, we explore the sweet spots with different relative strengths of ECT and CAR by tuning the on-site energies of superconducting dots.

*The Kitaev Chain Model.* To gain intuition on AKCs, let us first begin with a three-site Kitaev chain whose Hamiltonian can be written as:

$$H_{\mathrm{KC}} = \sum_{j=1}^{3} \varepsilon_{j,\mathrm{K}} n_j + \sum_{j=1}^{2}\left(t_{j,\mathrm{K}} c_j^\dagger c_{j+1} + \delta_{j,\mathrm{K}} e^{i\varphi_j} c_j^\dagger c_{j+1}^\dagger + \mathrm{h.c.}\right), \tag{8}$$

where $\varepsilon_{j,\mathrm{K}}$ is the on-site energy, $t_{j,\mathrm{K}}$ and $\delta_{j,\mathrm{K}}$ represents the strengths of ECT and CAR, respectively. $\varphi_j$ represents the phase of CAR process. We choose $\varphi_1 = 0$ and set $\varepsilon_{j,\mathrm{K}} = \varepsilon_{\mathrm{K}}$, $t_{j,\mathrm{K}} = t_{\mathrm{K}}$, $\delta_{j,\mathrm{K}} = \delta_{\mathrm{K}}$ and $\varphi_2 = 0$ unless otherwise stated. Figure 4(a) shows the zero-energy solutions corresponding to negative values of $t_K/\delta_K$, where three arrows (blue, green and red) mark three kinds of sweet spots in three parameter regimes: $|t_K| > |\delta_K|$, $|t_K| = |\delta_K|$ and $|t_K| < |\delta_K|$. Notably, despite the variation in $t_K/\delta_K$, the degenerate ground states persist at $\varepsilon_K = 0$. To illustration the origin of this robustness, we rewrite the Hamiltonian terms of Majorana operators at $\varepsilon_K = 0$, by introducing $c_j^\dagger = (\gamma_{j,A} + i\gamma_{j,B})/2$:

$$H_{KC} = -\frac{i}{2}\sum_{j=1}^{2}\left[(t_K - \delta_K)\gamma_{j,A}\gamma_{j+1,B} + (t_K + \delta_K)\gamma_{j+1,A}\gamma_{j,B}\right], \tag{9}$$

with the coupling sketch shown in Fig. 4(b). When $|t_K| = |\delta_K|$, the $\gamma_{1B}$ and $\gamma_{3A}$ (or $\gamma_{1A}$ and $\gamma_{3B}$) modes are well localized at the ends. When $|t_K| \neq |\delta_K|$, the $\gamma_{1A}$, $\gamma_{2B}$ and $\gamma_{3A}$ (as well as $\gamma_{1B}$, $\gamma_{2A}$ and $\gamma_{3B}$) combine to form a conventional fermion, leaving a Majorana bound state uncoupled. This distinction is illustrated by the Majorana wavefunctions shown in Fig. 4(c). In the left panel (green), the $\gamma_A$ and $\gamma_B$ are fully localized at the right and the left ends, respectively. In the right panel (red), small components of $\gamma_A$ and $\gamma_B$ leak into the opposite ends. In the following analysis, we classify the sweet spots based on the relative strengths of ECT and CAR: the ECT-dominated sweet spots with $|t_K| > |\delta_K|$, the genuine sweet spots with $|t_K| = |\delta_K|$, and the CAR-dominated sweet spots with $|t_K| < |\delta_K|$.

We investigate the energy-phase relations at these sweet spots. Figures 4(d), (e) and (f) show the energy differences $\delta E_{mn}^{\alpha\beta}$ as functions of the phase difference $\varphi = \varphi_2 - \varphi_1$, corresponding to the blue, green and red arrows in Fig. 4(a), respectively. As shown, $\delta E_{00}^{oe}$ remains zero throughout the variation of $\varphi$ at all these sweet spots, indicating the degenerate ground states. $E_m^o$ and $E_m^e$ also maintain their degeneracy regardless the variation of $\varphi$. $E_1^o$ and $E_2^o$ are degenerate at $\varphi = 0$ with the excitation gap $\delta E_{gap} = \sqrt{2(t^2 + \Delta^2)}$. $E_1^o$ gradually decreases as $\phi$ sweeps from 0 to π, resulting in $\delta E_{gap} = \sqrt{2}|t - \Delta|$ at $\varphi = π$. Therefore, $\delta E_{gap}$ vanishes at the genuine sweet spot when $\varphi = π$, but remains finite at both the ECT-dominated and CAR-dominated sweet spots. This finite excitation gap regardless of $\varphi$ is particularly helpful for cQED measurements. Firstly, the absence of such a gap will induce unavoidable energy-level crossings between plasma modes and the excited states of the chain, leading to signal interferences. Secondly, the presence of such a gap can suppress the probability of Landau-Zener transitions from the ground state to the excited states during rapid phase manipulations.

*The AKC Model*. Let us revisit the three-site AKC as shown in Fig. 5(a). Under the assumption of left-right symmetry, our analysis focuses on the regime near the genuine sweet spot, with small deviations $\delta\varepsilon_2 = \delta\varepsilon_4 < 0.2\Delta$. The on-site energies of quantum dots $\varepsilon_{j,SP}$ ($j = 1,3,5$) are optimized to match the characteristic in Figs. 4(d)-(f), and their values are summarized in Fig. 5(b). All sweet spots have high Majorana polarizations at ends of the chain, $|MP_1| > 0.95$, with the genuine sweet spot displaying the highest polarization value. To understand the correspondence between Fig. 5(b) and Fig. 4(a), we calculate the strengths of ECT and CAR between dot 1 and dot 3, denoted by $\Gamma_{\eta\sigma}^{ECT}$ and $\Gamma_{\eta\sigma}^{CAR}$ respectively. Figure 5(c) only shows $|\Gamma_{\downarrow\downarrow}^{ECT}|$ and $|\Gamma_{\downarrow\downarrow}^{CAR}|$, because the spin-↓ electrons dominate the chain's behavior in the presence of Zeeman field. As $\delta\varepsilon_{2,4}$ sweeps from $0.2\Delta$ to $-0.2\Delta$, $|\Gamma_{\downarrow\downarrow}^{ECT}|$ and $|\Gamma_{\downarrow\downarrow}^{CAR}|$ show a transition from CAR dominance to ECT dominance, with equal strengths at $\delta\varepsilon_2 = \delta\varepsilon_4 \approx 0.06\Delta$. The small deviation may arise from the finite Zeeman splitting.

Similar to Fig. 4(a), we use blue, green, and red arrows to mark three kinds of sweet spots in Fig. 5(b), respectively. Their energy-phase relations are shown in Figs. 5(d)-(f). These results are qualitatively similar to those in Figs. 4(d)-(f), particularly the non-vanishing excitation gaps $\delta E_{\text{gap}}$ at both ECT-dominated (blue) and CAR-dominated (red) sweet spots, whereas the gap nearly vanishes at the genuine sweet spot. The energy differences between the excited states and the ground state with same parity, $\delta E_{m0}^{\alpha\alpha}$ ($m > 0$), can be directly detected via cQED measurements. Therefore, the corresponding curves in Figs. 5(d)-(f) arise from the excitation spectrum of the chain. We can also distinguish the genuine sweet spot from the other two types by measuring $\delta E_{\text{gap}}$ at $\phi = \pi$. Moreover, Fig. 5(g) shows the Majorana wavefunctions at the green arrow (left panel) and the red arrow (right panel). Except for the small leakage into the superconducting dots, the behaviors of $\gamma_A$ and $\gamma_B$ agree with those in Fig. 4(c).

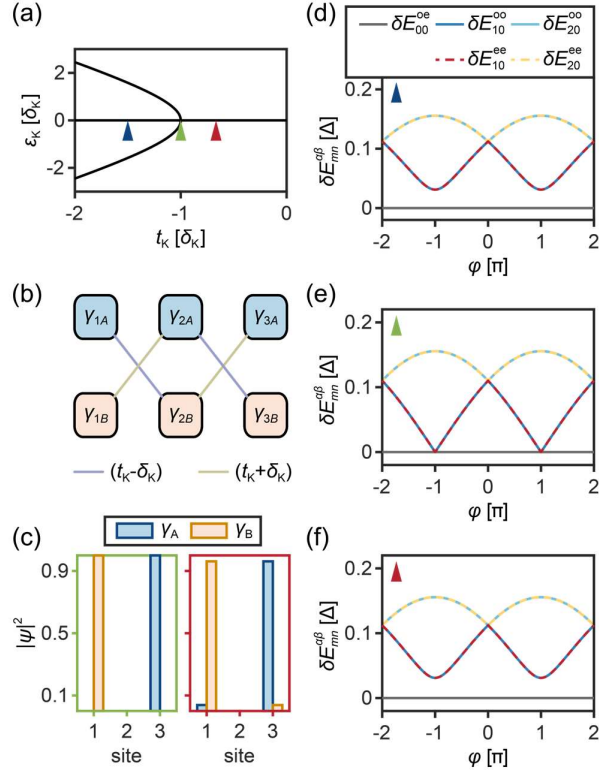

FIG. 4. (a) The zero-energy solutions of the three-site Kitaev chain with negative values of $t_K/\delta_K$ (reproduced from Ref. [16]). Blue, green and red arrows mark positions where $t_K/\delta_K = -1.5, -1.0$ and $-0.67$, respectively, with $\varepsilon_K = 0$. (b) Schematic illustration of the couplings in the Kitaev model at $\varphi = 0$. (c) The Majorana wavefunctions at the green and red arrows in the left and right panels, respectively. (d-f) $\delta E_{mn}^{\alpha\beta}$ as functions of the phase difference $\varphi$ at the blue (the ECT-dominated sweet spot), green (the genuine sweet spot) and red arrows (the CAR-dominated sweet spot). The parameters $(t_K, \delta_K)$ marked by these arrows are set to $(-1.2, 0.8)\delta_0$, $(-1.0, 1.0)\delta_0$, $(-0.8, 1.2)\delta_0$, respectively, where $\delta_0 = 0.055\,\Delta$.

Finally, we discuss how phase manipulations can probe the degenerate ground states at the sweet spots. In contrast to Figs. 4(d)-(f), the energy splitting $\delta E_{00}^{oe}$ in Fig. 5(d)-(f) does not remain constant, but exhibit slight variations with $\phi$, particularly near $\phi = \pi$. These results are consistent with those reported in Ref. [20]. Notably, this provides a phase window to determine the number of degenerate states. Figures 5 (h) and (i) show $\partial_\phi^2 E_0^{o/e}$ at the green and red arrows, respectively. At the genuine sweet spot, $\partial_\phi^2 E_0^o$ and $\partial_\phi^2 E_0^e$ remain coincident across nearly the entire phase range, splitting only within an extremely narrow vicinity of $\phi = \pi$ where sharp dips occur in both quantities. At the CAR-dominated sweet spot, the splitting exists across a broader range, but displays a smaller magnitude at $\phi = \pi$. A larger splitting improves the distinguishability in the transmon spectrum, whereas a broader phase range reduces the difficulty of phase control by an external magnetic flux. A balance between these two factors should be considered in experiments.

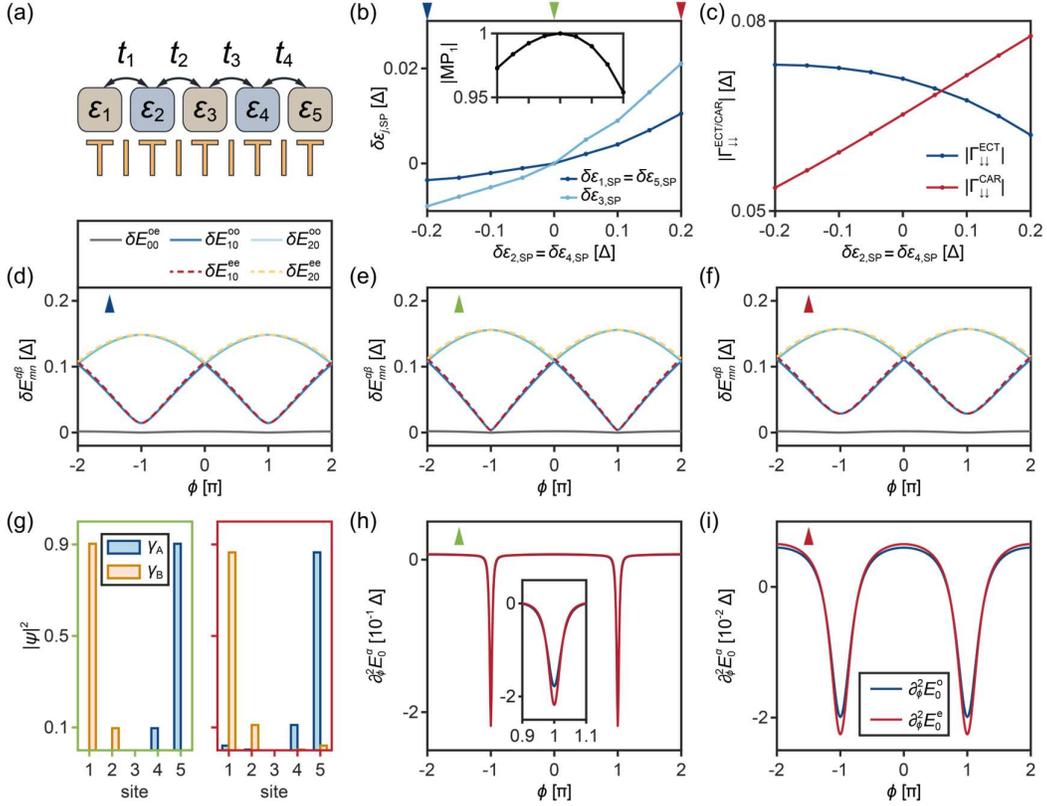

**Fig. 5.** (a) Schematic illustration of the QD-S-QD-S-QD configuration, i.e. a fully coupled three-site AKC. (b) $\delta\varepsilon_{j,\mathrm{SP}}$ ($j = 1,3,5$) as functions of $\delta\varepsilon_{(2,4),\mathrm{SP}}$ near the genuine sweet spot. The inset shows the |MP| at dot 1. At the genuine sweet spot, $\varepsilon_{1,\mathrm{SP}} = \varepsilon_{5,\mathrm{SP}} \approx 2.915\Delta$, $\varepsilon_{2,\mathrm{SP}} = \varepsilon_{4,\mathrm{SP}} \approx -0.695\Delta$ and $\varepsilon_{3,\mathrm{SP}} \approx 2.847\Delta$. The blue, green and red arrows mark $\delta\varepsilon_{2,\mathrm{SP}} = -0.2\Delta, 0$, and $0.2\Delta$, respectively. (c) $|\Gamma_{\downarrow\downarrow}^{\mathrm{ECT}}|$ and $|\Gamma_{\downarrow\downarrow}^{\mathrm{CAR}}|$ between dot 1 and dot 3 as functions of $\varepsilon_{2,4}$, using the parameters in panel (b). (d-f) $\delta E_{mn}^{\alpha\beta}$ as functions of the phase difference $\varphi$ at the blue (the ECT-dominated sweet spot), green (the genuine sweet spot) and red arrows (the CAR-dominated sweet spot), respectively. (g) The Majorana wavefunctions at the green and red arrows in the left and right panels, respectively. (h) and (i) $\partial_\phi^2 E_0^{o/e}$ as functions of $\phi$ at the genuine sweet spot and the CAR-dominated sweet spot. The inset in (h) shows a magnified view near $\phi = \pi$. The splitting at $\phi = \pi$ is approximately $4 \times 10^{-2}\Delta$ in (h) and $3 \times 10^{-3}\Delta$ in (i).

## V. Conclusion

In this work, we propose a procedure for tuning up a three-site AKC to its sweep spots based on the quantum electrodynamic spectra of a transmon circuit. Our approach enables fast and non-destructive measurements of AKC's quantum states. We have shown that there are three types of sweet spots, ECT-dominated sweet spots, genuine sweet spots and CAR-dominated sweet spots. And our analysis demonstrates that the ECT-dominated and CAR-dominated sweet spots can be more conveniently utilized in cQED measurements. In addition, we have investigated the detection of degenerate ground states under phase manipulations, and found that the degeneracy can be directly observed if the energy differences of these states are within the energy scale of thermal fluctuations. Our analysis reveals that in the QD-S-QD-S configuration the signatures of parity splitting across the entire phase range, but in the QD-S-QD-S-QD configuration the splitting is confined to a narrower region near $\phi = \pi$. We note that direct observation of degenerate ground states at sweet spots has not been reported in previous work, and hope that the observation of degeneracy can provide additional evidence for the MBSs.

## Acknowledgement

We thank Chun-Xiao Liu for valuable comments and discussions. This work was supported by the Innovation Program for Quantum Science and Technology through Grant No. 2021ZD0302600; by the National Natural Science Foundation of China (NSFC) through Grant Nos. 92365207, 92065203, 92365302, 11527806, 12074417, 11874406, 11774405 and E2J1141; by the Strategic Priority Research Program B of the Chinese Academy of Sciences through Grants Nos. XDB33010300, DB28000000, and XDB07010100; by the National Basic Research Program of China through MOST Grant Nos. 2016YFA0300601, 2017YFA0304700 and 2015CB921402; by Beijing Natural Science Foundation through Grant No. JQ23022; by Beijing Nova Program through Grant No. Z211100002121144; and by Synergetic Extreme Condition User Facility (SECUF).